\documentclass[]{aastex631}

\usepackage{CJK}
\usepackage{booktabs}
\usepackage{threeparttable}
\usepackage{amsmath}
\usepackage[graphicx]{realboxes}
\usepackage{enumerate}
\usepackage{overpic}
\usepackage{xcolor} 
\usepackage{pict2e} 

\usepackage{soul}

\shorttitle{First Detection of A Radio Nuclear Ring and Potential LLAGN in NGC 5792}
\shortauthors{Yang Yang et al.}
\graphicspath{{./}{figures/}}

\begin{document}
\begin{CJK*}{UTF8}{gbsn}

\title{CHANG-ES. XXIV. First Detection of A Radio Nuclear Ring and Potential LLAGN in NGC 5792}

\correspondingauthor{Yang Yang}
\email{yangyang.astro@gmail.com}

\author[0000-0001-7254-219X]{Yang Yang (杨阳)}
\affiliation{Shanghai Astronomical Observatory, Chinese Academy of Sciences, 80 Nandan Road, Shanghai 200030, People’s Republic of China}

\author[0000-0003-0073-0903]{Judith Irwin}
\affiliation{Department of Physics, Engineering Physics \& Astronomy, Queens University, Kingston, Ontario, K7L 3N6, Canada}

\author[0000-0001-6239-3821]{Jiangtao Li}
\affiliation{Department of Astronomy, University of Michigan, 311 West Hall, 1085 South University Avenue,  Ann Arbor, Michigan, 48109, USA}

\author[0000-0002-3502-4833]{Theresa Wiegert}
\affiliation{Department of Physics, Engineering Physics \& Astronomy, Queens University, Kingston, Ontario, K7L 3N6, Canada}

\author[0000-0002-9279-4041]{Q. Daniel Wang}
\affiliation{Department of Astronomy, University of Massachusetts, LGRT-B 619E, 710 North Pleasant Street, Amherst, MA 01003-9305, USA} 

\author[0000-0002-5456-0447]{Wei Sun}
\affiliation{Purple Mountain Observatory, Chinese Academy of Sciences, 10 Yuanhua Road, Nanjing, Jiansu 210093, People’s Republic of China}

\author{A. Damas-Segovia}
\affiliation{MPI f{\"u}r Radioastronomie, Auf dem H{\"u}gel 69, 53121 Bonn, Germany}

\author{Zhiyuan Li}
\affiliation{School of Astronomy and Space Science, Nanjing University, Nanjing 210023, People’s Republic of China}
\affiliation{Key Laboratory of Modern Astronomy and Astrophysics, Nanjing University, Nanjing 210023, People’s Republic of China}

\author{Zhiqiang Shen}
\affiliation{Shanghai Astronomical Observatory, Chinese Academy of Sciences, 80 Nandan Road, Shanghai 200030, People’s Republic of China}
\affiliation{31 Key Laboratory of Radio Astronomy, Chinese Academy of Sciences, Nanjing 210008, People’s Republic of China}

\author[0000-0002-0782-3064]{Ren{\'e} A. M. Walterbos}
\affiliation{Department of Astronomy, New Mexico State University, Las Cruces, NM 88003, USA}

\author{Carlos J. Vargas}
\affiliation{Department of Astronomy and Steward Observatory, University of Arizona, Tucson, AZ, USA}


\begin{abstract}

We report the discoveries of a nuclear ring of diameter 10$\arcsec$ ($\sim$1.5 kpc) and a potential low luminosity active galactic nucleus (LLAGN) in the radio continuum emission map of the edge-on barred spiral galaxy NGC~5792. These discoveries are based on the Continuum Halos in Nearby Galaxies - an Expanded Very Large Array (VLA) Survey, as well as subsequent VLA observations of sub-arcsecond resolution. 
Using a mixture of H$\alpha$ and 24 $\mu$m calibration, we disentangle the thermal and non-thermal radio emission of the nuclear region, and derive a star formation rate (SFR) of $\sim 0.4~M_{\sun}$ yr$^{-1}$.  
We find that the nuclear ring is dominated by non-thermal synchrotron emission. The synchrotron-based SFR is about three times of the mixture-based SFR. This result indicates that the nuclear ring underwent more intense star-forming activity in the past, and now its star formation is in the low state. 
The sub-arcsecond VLA images resolve six individual knots on the nuclear ring.
The equipartition magnetic field strength $B_{\rm eq}$ of the knots varies from 77 to 88 $\mu$G. 
The radio ring surrounds a point-like faint radio core of $S_{\rm 6GHz}=(16\pm4)$ $\mu$Jy with polarized lobes at the center of NGC~5792, which suggests an LLAGN with an Eddington ratio $\sim10^{-5}$.
This radio nuclear ring is reminiscent of the Central Molecular Zone (CMZ) of the Galaxy. Both of them consist of a nuclear ring and LLAGN.

\end{abstract}


\keywords{galaxies: individual (NGC~5792) --- galaxies: nuclei --- galaxies: star formation --- radio continuum: galaxies}


\section{Introduction}  \label{sec:intro}

Edge-on galaxies are ideal laboratories for studying gas interchange between star-forming (SF) disks and the surrounding extraplanar and halo environment. Probing the properties of the medium located at the disk-halo interface allows us to study how the gas transfers into the disk, forms spiral arms, ring, and bar structures, as well as sustains the star-forming activity. The accumulation of in-flowing gas in the nuclear region provides the necessary material for the activities of starbursts (SB) and active galactic nucleus (AGN), and the environment of host galaxies is affected by the feedback of these activities. Therefore, the study of these feedback activities are essential to understanding the evolution of galaxies. The radio emission is free from extinction in the nuclear region, since high-spatial (sub-arcsecond) resolution radio observations of nuclei are a useful way to understand how SB and AGN activities affect the evolution of host galaxies. 



\begin{figure}
\centering
\begin{overpic}[width=1.0\textwidth]{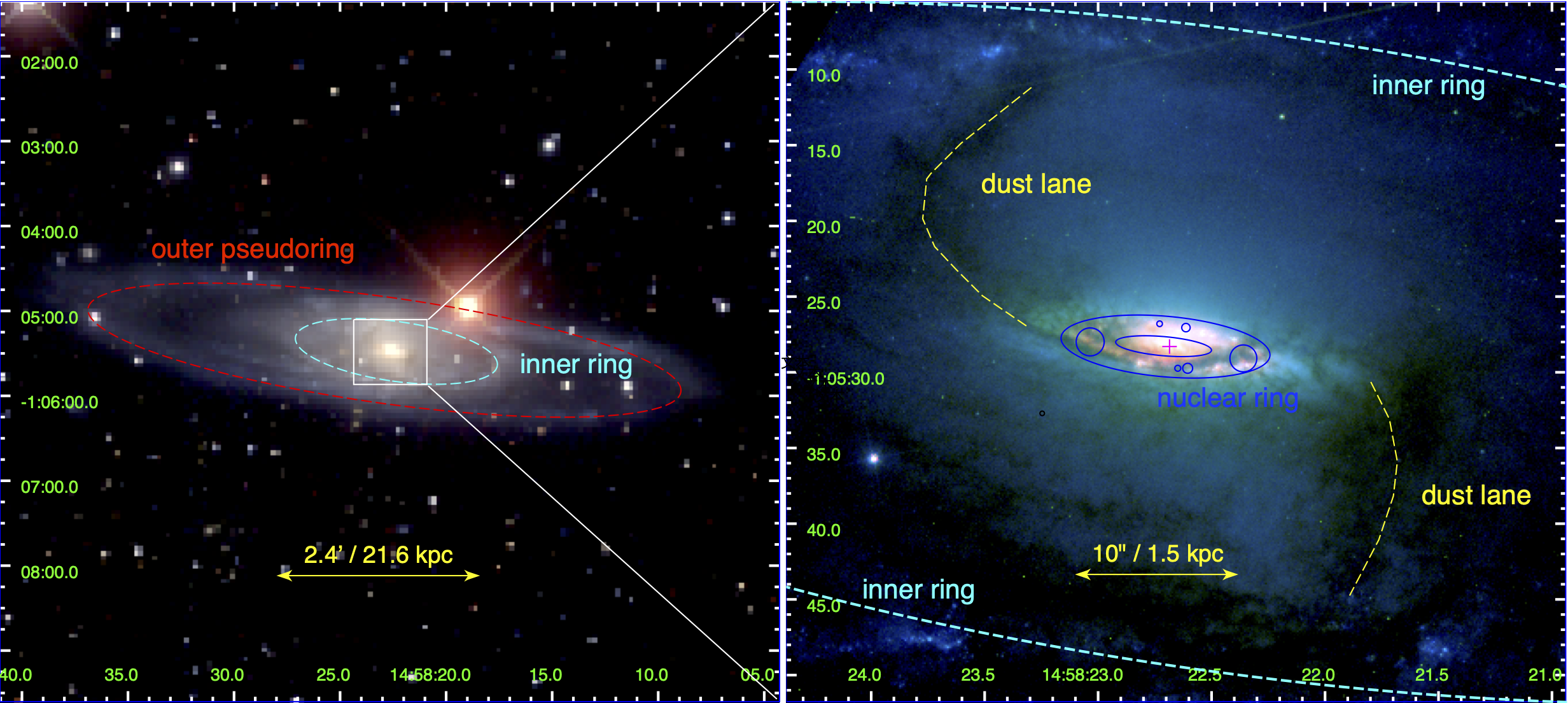}
\end{overpic}
\caption{\label{fig:sdss} Global Morphology. Left panel: color composite image of the global morphology of NGC 5792, created by stacking Sloan Digital Sky Survey (SDSS) images in g, r, and i filters. The g filter is shown in blue, while r and i are shown in green and red, respectively. The red ellipse represents the outer pseudo-ring, and the cyan ellipse represents the inner ring. 
Right panel: A blow-up of the the white rectangular region in the left panel, with the main dynamical features overlaid on an \emph{HST} map (same as in figure \ref{fig:Xband}d)). The cyan line indicates (part of) the inner ring and the dashed yellow arcs represent the dust lanes.}
\end{figure}

NGC~5792 is a nearby, highly inclined ($i\sim70-80\degr$), barred spiral galaxy at a distance of about 31.7 Mpc (1$\arcsec\approx$ 150 pc)
in the constellation Libra \citep{2011AA...532A..74B, 2012AJ....144...43I}.
Previous optical studies indicate that NGC~5792 has an outer pseudo-ring of radius 31.5 kpc and an inner ring of radius 10.8 kpc \citep[][see also Figure \ref{fig:sdss}]{2014AA...562A.121C}. Some basic parameters of NGC~5792 are listed in Table \ref{tab:galaxy}.

Recently, several authors have suggested that the nuclear ring of barred galaxies are regions of large gas surface densities and high star formation rates \citep{1996FCPh...17...95B, 2005A&A...429..141K, 2010MNRAS.402.2462C}. Nuclear rings are believed to form as a result of gas inflow toward the central region along dust lanes, where it stagnates near dynamical resonances.  
Due to the bar torque, the in-flowing material, which loses most of its angular momentum, spirals in toward the ring region at the two ``contact points" between the dust lanes and the ring, where the accumulated gas proceeds to move in nearly circular orbits and form a luminous, compact ring around the galactic center \citep{1992MNRAS.259..345A, 1994ApJ...424...84H, 1996ApJ...471..143H, 1995ApJ...449..508P, 1996FCPh...17...95B, 1997ApJ...482L.143R, 2002AJ....123.1411B,2012ApJ...747...60K,2015ApJ...806..150L,2018ApJ...857..116M}. 
Typically located within the central kiloparsec, nuclear rings contain a mixture of neutral, ionized gas, and dust with total masses of 10$^8-10^{10}$ $M_{\sun}$  \citep{1996ApJ...471..143H, 1997AJ....113.1250R}. One of the most well-studied nuclear rings is the region of abundant molecular gas in our Milky Way galaxy, called the Central Molecular Zone (CMZ).

Our new Karl G. Jansky Very Large Array (VLA) A-configuration observations provide sub-arcsecond resolving power in the radio regime, allowing us to probe the nuclear region in NGC~5792. The goal of this paper is to provide an overview of the nuclear ring and to characterize its nature, on the basis of our high-resolution multi-wavelength data, as well as to compare the nuclear ring with the CMZ in our Galaxy. 
Furthermore, we aim to probe the possible existence of an AGN at the center of NGC 5792.

In this paper, we present the results of Continuum Halos in Nearby Galaxies - an EVLA Survey (CHANG-ES) observations \citep[][]{2012AJ....144...43I} for NGC~5792, and use our follow-up sub-arcsecond angular resolution VLA (6 and 9 GHz) observations, near-IR (Hubble Space Telescope (HST) WFC3 F160W and F814W filter), and optical (H$\alpha$, \emph{HST}/WFC3 F475W) images to investigate the central kiloparsec region of NGC~5792 in detail. 
We present an overview of the observations in \S2. The information and morphological properties of our results are presented in \S3. Our detailed analysis of the results is discussed in \S4, with concluding remarks in \S5.

\begin{table*}
\caption{Basic galaxy parameters.}
\vspace{0.2cm}
{
\begin{tabular}{ll}
\hline
\hline
\multicolumn{2}{c}{NGC 5792 (PGC0053499 )}\\
\hline
RA (J2000)&14$^{\rm h}$58$^{\rm m}$22.71$^{\rm s}$\\
Dec (J2000)&-01\arcdeg05\arcmin27.9\arcsec\\
Distance (Mpc)&31.7 \citep{2012AJ....144...43I}\\
Inclination of disks  ($\degr$)&70-80 \citep{2011AA...532A..74B}\\
PA of major axis$^a$ ($\degr$)&84 \citep{2004AA...415..849B}\\ 
Classification&SBb\\
Nuclear Type&HII\\
SFR$^b$ ($M_{\sun}$ yr$^{-1}$)   &4.41 \citep{2019ApJ...881...26V}\\
M$_\bullet^c$ ($M_{\sun}$)& 10$^{7.17}$ \citep{2014ApJ...789..124D}\\
M$_{\star}^d$ ($M_{\sun}$) &10$^{10.93 \pm 0.01}$ \citep{2019PASJ...71S..14S} \\
M$_{gas}^e$ ($M_{\sun}$) &10$^{10.31}$ \citep{2004AA...417...39M} \\
M$_{dust}^f$ ($M_{\sun}$) &10$^{7.78}$ \citep{2004AA...417...39M} \\
Diameter of the outer pseudo-ring ( \arcsec/ kpc, 1$\arcsec\approx$ 150 pc)&413/63 \citep{2014AA...562A.121C} \\
Diameter of the inner ring ( \arcsec/ kpc, 1$\arcsec\approx$ 150 pc)&146/22 \citep{2014AA...562A.121C} \\
\hline
\end{tabular}}
\vspace{0.2cm}

\textbf{Notes}:$^a$ The position angle (PA) is the angle between the line of nodes of the projected image and the north, measured towards the east. $^b$ Star Formation Rate (SFR) in 23.7 kpc diamater. $^c$ The mass of the supermassive black hole. $^d$ Stellar mass in 72 kpc diamater. $^e$ The total gas mass of this galaxy. $^f$ The dust mass of this galaxy.
\label{tab:galaxy}
\end{table*}

\section{Observations Data Reductions}   \label{sec:obs}
    \subsection{CHANG-ES Observations}
    NGC 5792 is one of the 35 highly-inclined nearby galaxies in the CHANG-ES project. The CHANG-ES observations were carried out using the updated VLA during its commissioning period (2011-2012) at L-band (center frequency at 1.5 GHz, in B, C, and D configurations) and C-band (center frequency at 6 GHz in the C and D configurations), in all polarization products. 
    Details of the CHANG-ES observations for NGC~5792 are listed in Table \ref{tab:obs}.

    \subsection{High-spatial (sub-arcsecond) Resolution VLA Observations}
    In addition to the CHANG-ES observations, we observed the nuclear region of NGC~5792 in 3 epochs (project ID: 15A-400; PI: Yang) with dual-polarization with the VLA in A-configuration. These data were observed at C-band (centered at 6 GHz) and X-band (centered at 9 GHz). Our observations were centered at [R.A., Dec] (J2000) = [14$^{\rm h}$58$^{\rm m}$22.71$^{\rm s}$, -01\arcdeg05\arcmin27.9\arcsec]. Corresponding observation logs are summarized in Table \ref{tab:obs}. We used the same flux calibrator and secondary calibrator as in the CHANG-ES observations.
    
   \subsection{Data Reductions}
   Each individual visibility data were flagged, calibrated, imaged and restored using the Common Astronomy Software Applications package (CASA, version 4.5) following the standard procedures  \citep[more details can be found in][\S2.2]{2012AJ....144...44I}. We inspected all visibility data by \textbf{eye}, and manually flagged bad data (due to radio frequency interference and instrumental effects). The Stokes I images were then produced using the \texttt{CLEAN} task, with the Multi-frequency Synthesis mode, nterms = 2, and the Briggs weighting with robust = 0. The CASA {\texttt{WIDEBANDPBCOR}} task was used to carry out wide-band primary beam corrections. Flux measurements were made from the primary beam-corrected images in all images. 
   To estimate the radio spectral index between 6 and 9 GHz associated with each region. We smoothed the native resolution 6 GHz and 9 GHz images to a common circular beam (FWHM = 0.3\arcsec) for consistent measurements of flux densities across the two frequencies. All flux densities (listed in Col. 5 and Col. 6 of Table \ref{tab:flux}) are measured by using {\texttt{IMSTAT}} task in the corresponding region, whose the position and radius are listed in Col. 2 and Col. 3 of Table \ref{tab:flux}, and we will state how to define the corresponding regions in \S \ref{sec:mophy}.
   The root-mean-square error (RMS) is measured in a signal-free portion (near-source) of each image. The uncertainty of the flux densities for each region was derived via the equation
$\sigma\simeq\sqrt{N_b\times RMS^2 + (\eta\times S)^2}$, in which $N_b$ corresponds to the number of synthesized beams, and $\eta$ is a factor that accounts for uncertainties in the calibration system, which we adopted $\eta$ = 0.03 for the VLA radio images \citep{2013ApJS..204...19P}; $S$ is the flux density of the corresponding region. Only the flux density of the core was measured by fitting a Gaussian to the corresponding region in the 6 GHz A-array image with the \texttt{IMFIT} task, its fitting region is approximately 2 times the full-width-half-maximum of the synthesized bea.
To maximize the signal-to-noise ratio, we used the \texttt{CONCAT} task to combine the two-epoch visibility data sets (at X-band in A configuration) into a concatenated visibility data set. 

\begin{figure}
\centering
\begin{overpic}[scale=1.0, trim=30 210 20 140, clip]{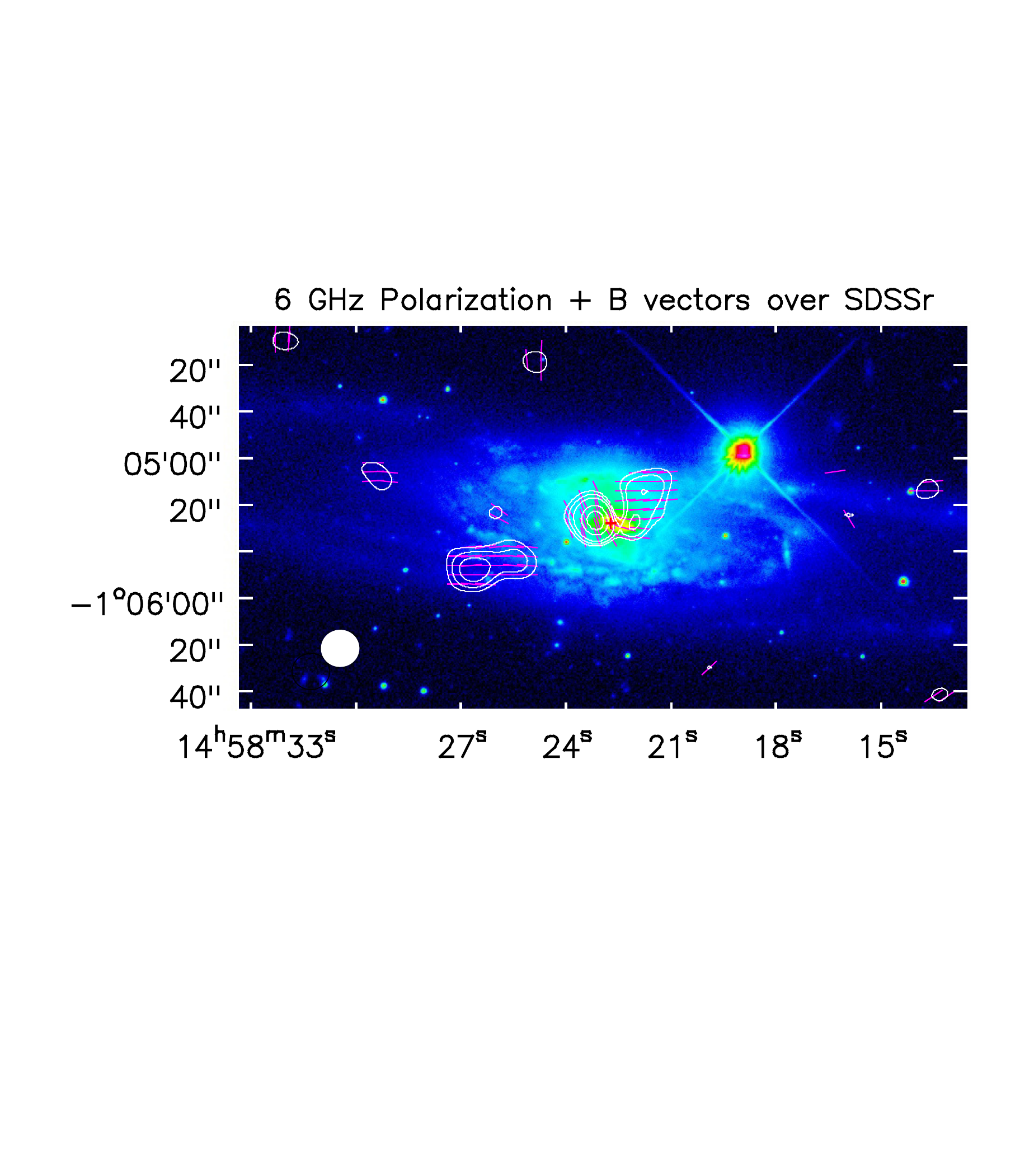}
\end{overpic}
\includegraphics[scale=0.72,  trim = -110 0 0 2]{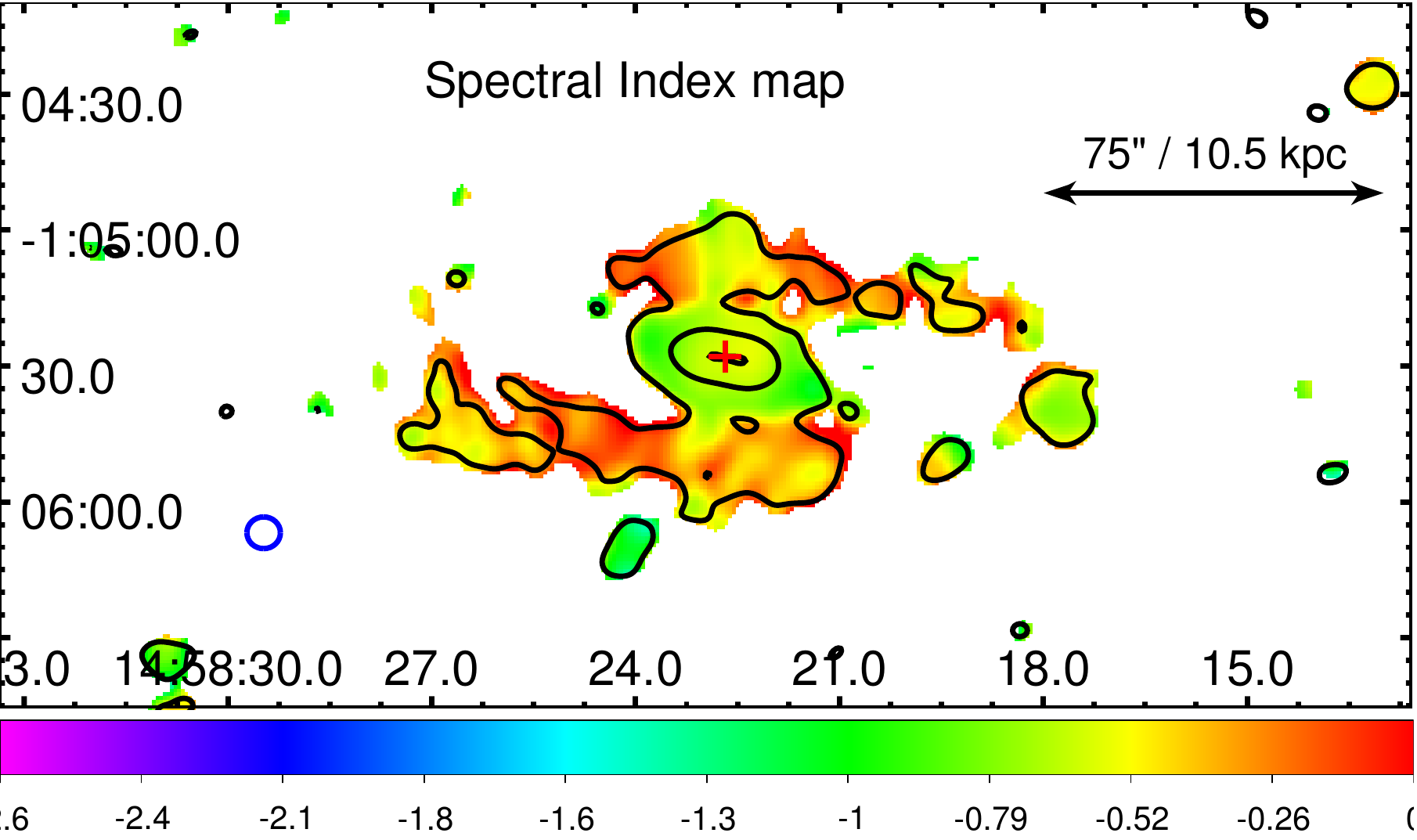}
\caption{Top: Polarization (P) contours and magnetic field  (B)  vectors overlaid on an optical SDSS r-band image.  The contours are at 16.2 (3$\sigma$), 22, 30, 50, and 70 $\mu$Jy beam$^{-1}$, where the beam size is shown in the lower left and has dimensions, $16.1^{\prime\prime} \times 15.5^{\prime\prime}$, $-87.4^\circ$. The B-vectors have been cut off at 5$\sigma$. The radio data are at 6 GHz and were observed with the VLA in D-configuration with a 6 k$\lambda$ uv taper applied during  imaging. Bottom: The spectral index distribution between the 6 GHz C-configuration data and 1.5 GHz B-configuration data. The angular resolution (blue ellipse) is $7.5\arcsec \times 6.7\arcsec$. Total intensity contour levels (1.5 GHz in B-configuration) are shown at an RMS of 20 $\mu$Jy beam$^{-1}$ $\times$ [3, 75, 435]. These two images show the same field size as the panel a) in Figure \ref{fig:CLband}.  The red cross points out the center of NGC~5792.}
\label{fig:pol}\label{fig:spix}
\end{figure}

Stokes Q and U maps (only in CHANG-ES observations) were formed with the same sets of input parameters as the total intensity images. 
We derived the linearly polarized intensity image as the top plane in Figure~\ref{fig:pol} by the relation of P=$\sqrt{Q^2+U^2-\sigma_{\rm Q,U}^2}$, where $\sigma_{\rm Q,U}$ is the RMS noise in the Q and U maps. The polarization angle of the observed electric vector ($\chi$) is given by $\chi=1/2\arctan(\rm U/Q)$; and the perpendicular direction of $\chi$ represents the apparent magnetic field orientation on the sky plane. However, the potential Faraday rotation is  uncorrected yet.

We also made the band-to-band spectral index map, as shown in Figure~\ref{fig:spix}, based on the 1.5-GHz B-configuration image as well as the 6.0-GHz C-configuration image of similar resolution.
We also have corrected the band-to-band spectral index map for the primary beam (PB) following the VLA Memo 195\footnote{https://library.nrao.edu/public/memos/evla/EVLAM\_195.pdf}. 
 
    \subsection{Archival H$\alpha$ and \emph{HST} Images}
In order to evaluate the star-forming and nuclear activities in the nuclear region, we used H$\alpha$ data \citep{2019ApJ...881...26V} observed by the Apache Point Observatory (APO) 3.5-m Telescope with the Astrophysical Research Consortium Telescope Imaging Camera (ARCTIC). The seeing-limited FWHM of the H$\alpha$ image is $\sim1.3\arcsec$, while
the pixel size is 0.228$\arcsec$ in Figure \ref{fig:Xband}c.
The reduction of the H$\alpha$ image is described in detail by \citet{2019ApJ...881...26V},
which includes bias removal, flat field correction, continuum and background subtraction, flux
calibration, as well as cosmic ray and other artifact removal.
The [\ion{N}{2}]-line contamination ($\sim0.15$) is small and was not corrected for. The uncertainty of the H$\alpha$ measured fluxes is taken to be $\sim 20$ \%.

In order to probe the stellar distribution and extinction in the nuclear region, we employed the high-resolution near-infrared/optical images from the \emph{HST} Legacy Archive (HLA, project ID: 15323) observed with the WFC3/UVIS and WFC3/IR instruments. A multi-color \emph{HST} image is shown as Figure \ref{fig:Xband}d.
The slight offset ($\sim0.3\arcsec$) between the centriod of radio core and the nucleus in the \emph{HST}/WFC3 image is consistent within the astrometric accuracy\footnote{\citealt{1998ApJ...502..199F} reported the pointing accuracy of both HST and VLA is 0.5$\arcsec$-2$\arcsec$. However, owing to the lack of common sources, it is virtually impossible to match the optical/radio astrometry to better than the individual pointing accuracy.}.

\section{Results}       \label{sec:res} 
    \subsection{The Morphology of the Radio Emission}   \label{sec:mophy}
    
\begin{figure}
\centering
\includegraphics[width=1.0\textwidth]{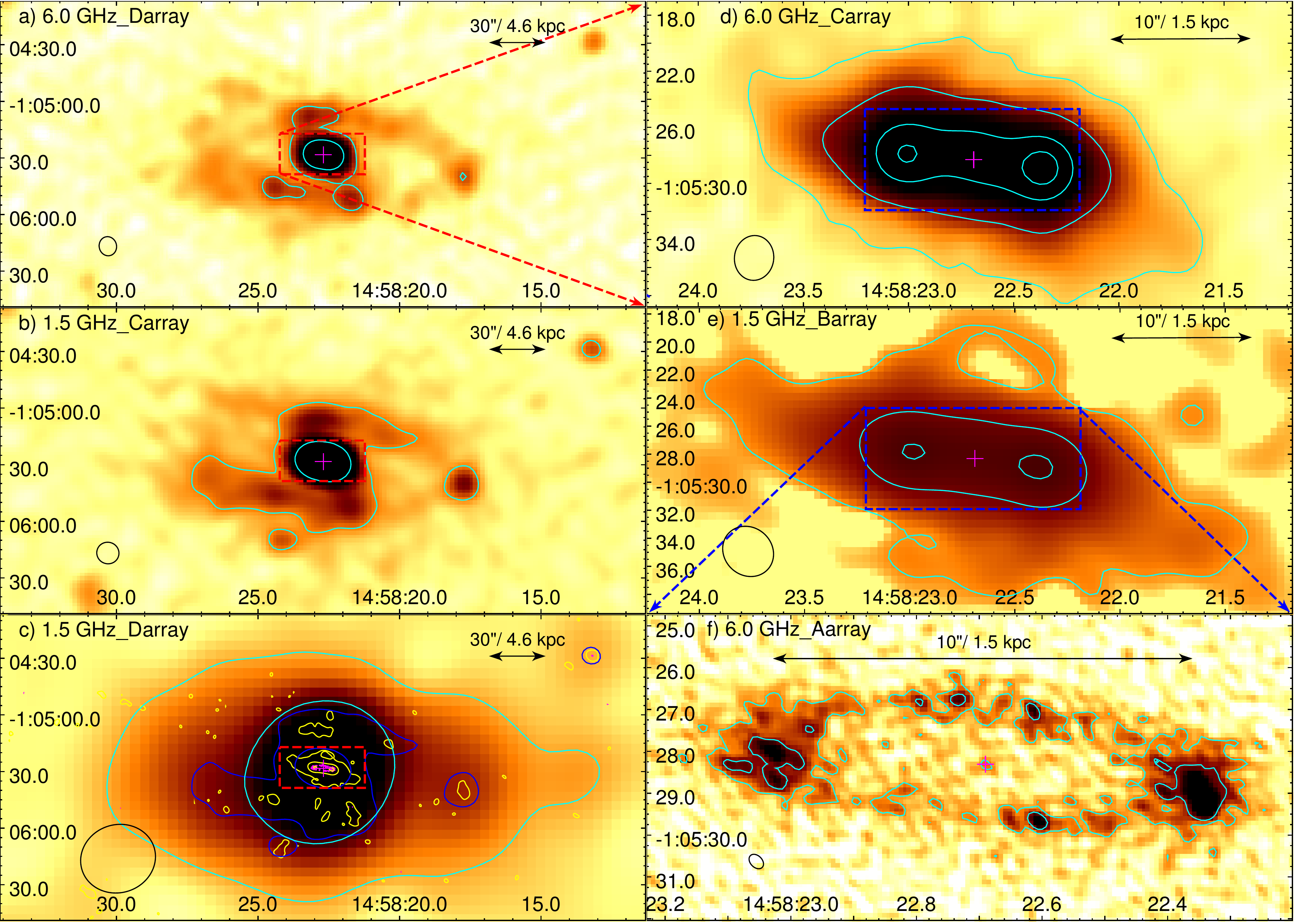}
\caption{\label{fig:CLband} Total radio intensity images with contours. 
The regions in a), b) and c) have the same scale. The red dashed box shown in these three planes is blown up to d) indicated by the arrows. The regions in d) and e) are of the same scale. The blue dashed box shown in these two planes is blown up to f) indicated by the arrows.
a) 6 GHz D-configuration image with contours at an RMS of 0.018~mJy~beam$^{-1}\times$[10, 100].
b) 1.5 GHz C-configuration image with contours at an RMS of 0.03 mJy~beam$^{-1}\times$[10, 100];
c) 1.5 GHz D-configuration image with cyan contours at an RMS of 0.045~mJy~beam$^{-1}\times$[10, 100], the blue contours show the corresponding contour from b), yellow contours show the corresponding contour from e), the magenta contours show the corresponding contour from f); 
d) 6 GHz C-configuration image with contours at an RMS of 0.018~mJy~beam$^{-1}\times$[1, 10, 50, 90];
e) 1.5 GHz B-configuration image with contours at an RMS of 0.02 mJy~beam$^{-1}\times$[3, 100, 225];
f) 6 GHz A-configuration image with contours at an RMS of 4~$\mu$Jy~beam$^{-1}\times$[3, 7];
The resolution (beam size) is indicated in the bottom left corner of each panel. The magenta cross indicates the position of the core. }
\end{figure}

\begin{figure}
\centering
\includegraphics[width=1.0\textwidth]{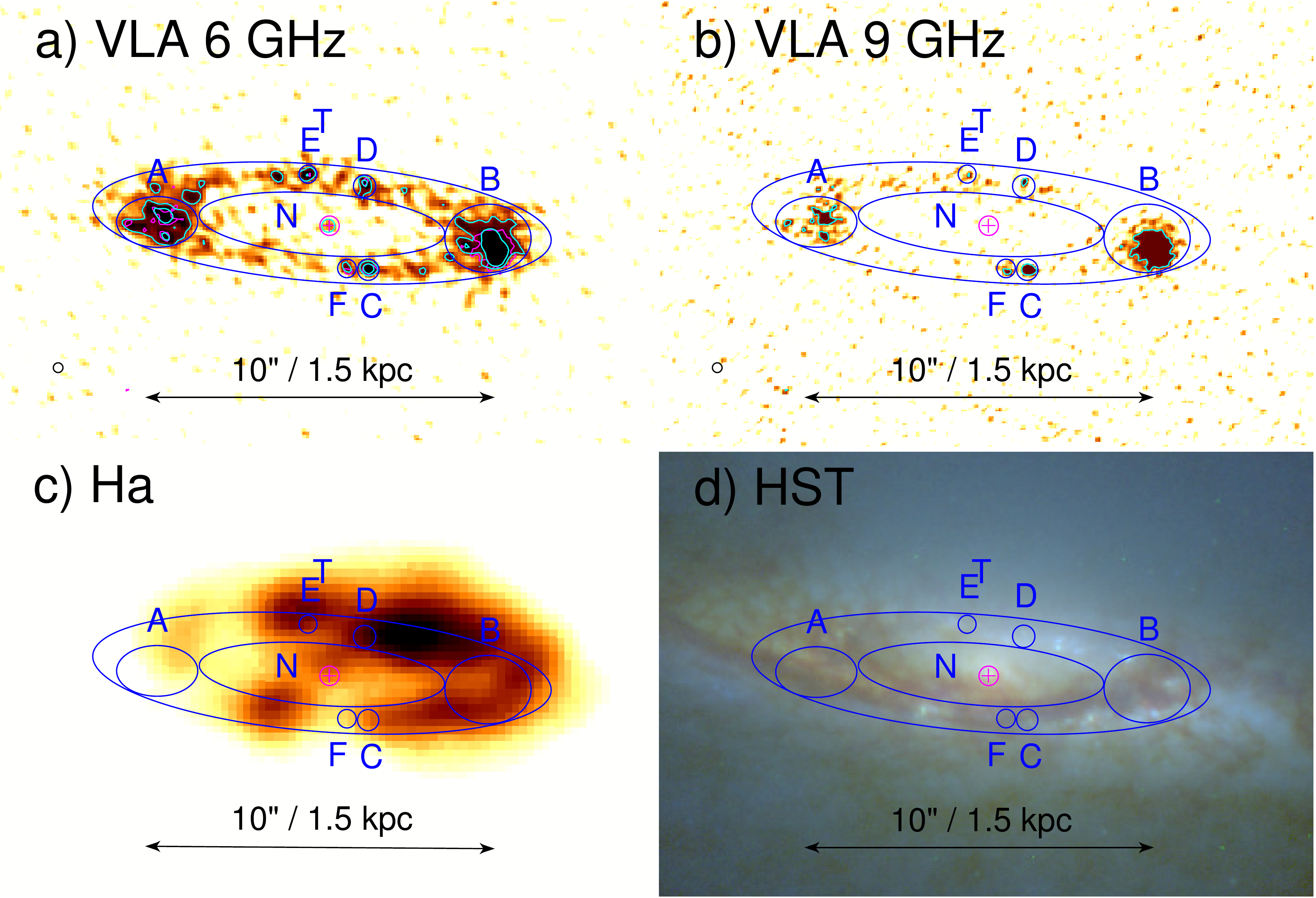}
\caption{a) Total intensity map from the 6 GHz A-configuration data, with contours at an RMS of 4~$\mu$Jy~beam$^{-1}\times$[6, 11],
b) Total intensity map from the 9 GHz A-configuration data, with contours at an RMS of 7~$\mu$Jy~beam$^{-1}\times$[3, 5],
c) The H$\alpha$ image (3.5m APO, filter: UNM 661/3), d) The \emph{HST} image (including F160W (H) filter in red, F814W (I) filter in green and F475W (g) in blue). 
We denote the inner region of the ring and the total nuclear region as N and T, respectively, and have labeled six knots A, B, C, D, E, and F on the ring. Region R is the ring defined by subtracting region N from the total nuclear region T. The region core is the point-like source marked as the magenta cross in the center of NGC~5792.}
\label{fig:Xband}
\end{figure}

   The 1.5 and 6 GHz total intensity images in different scales are shown in Figure~\ref{fig:CLband}, in which the panel a) - f) are arranged according to their field-of-views (FOVs) as well as frequencies descendingly and Figure~\ref{fig:CLband}a, b, and c are of the same FOV.
 Two arm-like structures extending from the nuclear region to a radius of $90\arcsec$ ($\sim{}14$~kpc) are discernable in the 6 GHz D-configuration and 1.5 GHz C-configuration maps (Figure~\ref{fig:CLband}a and b respectively). The two radio arm-like structures partially overlap the optical inner ring (see Figure \ref{fig:sdss} cyan ellipse). 
 The extended 1.5 GHz emission, as revealed in the D-configuration image (Figure~\ref{fig:CLband}c), appears to have a slight extension along the minor axis of the disk.

Besides that, in the 1.5 GHz B-configuration map (Figure \ref{fig:CLband}e), we detected a loop to the north of the nucleus with a signal to noise ratio $\sim3$, as well as two emission peaks in the nuclear region. Those two peaks are also visible in the 6 GHz C-configuration map (Figure~\ref{fig:CLband}d), and further resolved as a nuclear radio ring of diameter 1.5~kpc, with several knots spread over the ring, in the 6 (C-band) and 9 GHz (X-band) A-configuration maps (Figure~\ref{fig:Xband}a and b). We consider the regions within the 3-sigma contours to be from the real signals, and the 3-sigma contours at 9 GHz are just enclosed in the 6-sigma contours at 6 GHz. 
Therefore, we defined the knot regions where the intensity is 6 and 3 times above the RMS noise at 6 and 9 GHz respectively. The outermost cyan and magenta contours in Figure~\ref{fig:Xband}a represent the 6-sigma level at 6 GHz and 3-sigma level at 9 GHz, respectively. These knot regions are labeled as A, B, C, D, E, and F. The coordinates and sizes of the knots A to F are listed in Table~\ref{tab:flux}.

We also define the ellipse that just encloses these knots as the total nuclear region T and the inside ellipse just detached from these knots as region N. Region R is designated by subtracting region N from the total nuclear region T. 
The observed axis ratio of the nuclear ring, as measured in radio continuum, is 3.75, which implies an inclination of $i \sim 74.3^\circ\pm2.0^\circ$. This is consistent with the inclination (70-80 $\arcdeg$) of NGC 5792. 
In the center of the radio nuclear ring, we have detected a compact core with a flux density $16\pm4$ $\mu$Jy ($\sim 4\sigma$) at 6 GHz.

\subsection{Spectral Index of the Nuclear Ring}\label{sec:specind}

We made a band-to-band spectral index map from the 1.5 GHz B-configuration map to the 6 GHz C-configuration map of similar resolution ($7.5\arcsec\times6.7\arcsec$). The spectral index ($S\propto\nu^{\alpha}$) map was created using a cut-off at $3\sigma$ of the intensity maps \citep[see][\S3.4 for more details]{2019AJ....158...21I}. As shown in the Figure \ref{fig:spix}, the mean spectral index of the nuclear region is $\sim-0.7$ with the standard deviation of 0.15, 
while the mean spectral index of the north and south arms are $\sim-0.4$ with the standard deviation of 0.2 and $\sim-0.3$ with the standard deviation of 0.2. 
By setting the multiterm parameter nterms = 2, we also checked the mean in-band spectral index of the nuclear region, which is consistent with the band-to-band one.

In order to assess the spectral index in the nuclear ring, we used the higher-resolution ($\sim0.3\arcsec$) 6 and 9 GHz observations to derive the total spectral index $\alpha$ for each defined region in Figure \ref{fig:Xband}a and \ref{fig:Xband}b. The intensity maps of those two frequencies are convolved to a restoring beam of the same size (FWHM=0.3$\arcsec$) for consistency. The photometric information is summarized in Col. 5 and 6 in Table~\ref{tab:flux}, and derived total spectral index $\alpha$ is presented in Col. 7 of the same table. The $\alpha$ of the separate knots covers a range of $ -1$ -- $-3$, and that of the nuclear region as a whole is as low as $\sim -2.95\pm0.24$. 
The $\alpha_{6-9\rm~GHz}\sim-2.95\pm0.24$ is steeper than the $\alpha_{1.5-6\rm~GHz}\sim-0.70\pm0.15$ found by \citet{2016MNRAS.456.1723L}. 
Among the six knots, knot D has the steepest spectrum of $\alpha\sim -3.06\pm1.24$, while knot F has the flattest spectrum of $\alpha\sim -1.43\pm1.04$.
We also checked the in-band spectral index with the A-array data, but only got a meaningful value for the knot B at 6~GHz, as $-0.98\pm0.21$. This value is larger than the band-to-band spectral index (-1.65, see Table\ref{tab:flux}).
We note that region N and T are larger than the Largest Angular Scale (LAS), which may suffer  losing the 9~GHz flux density;
meanwhile, for the small knots whose flux density is as low as 3 times of RMS value, corresponding region sizes are close to the synthesized beam of the 9-GHz image, and may lose diffuse emission. 
As a consequence, we may have derived steeper spectra.
To assess the influence of the potential missing 9~GHz flux, we used observed 6-GHz flux density, adopted $\alpha_{1.5-6\rm~GHz}\sim-0.70$ to the 6--9~GHz range, and estimated the extrapolated $S_{\rm 9GHz,ext}$ in those regions. We found that those $S_{9GHz,ext}$ are about 1.3-2.5 times of the corresponding observed $S_{\rm 9GHz,obs}$.
Further physical discussion on the steep spectrum will be given in $\S\ref{sec:noninring}$.

We also detected a point-like source (marked as core) at the center of NGC~5792 in the 6 GHz image (Figure \ref{fig:CLband}f), with flux $\sim$ 4$\sigma$ above the average background level, but failed to detect in the 9 GHz map. We used its 3$\sigma$ upper limit to estimate the upper limit of the total spectral index of the core, and found a value of $\leq 0.66$. The non-thermal $\alpha_{NT}$ of the core is $\leq0.75$ after excluding the thermal contribution with the method described in \S\ref{sec:thermal}.

\subsection{Thermal (Free-Free) versus Non-Thermal (Synchrotron) Radio Emission} \label{sec:thermal}

The total observed radio emission is a combination of thermal free-free emission and non-thermal synchrotron emission, which have different origins \citep{1992ARA&A..30..575C}: massive stars and their associated \ion{H}{2} regions would be responsible for the thermal free-free  emission, while supernovae (SNe), supernova remnants (SNRs), and AGN 
would account for the non-thermal synchrotron emission \citep{2008A&A...477...95C,2017ApJ...843..117B}. It is necessary to separate these two kinds of emissions, in order to probe their origin and the processes behind them.

We estimated the thermal (free-free) radio emission contribution from extinction-corrected H$\alpha$ measurements. When galaxies are edge-on or very dusty,
\cite{2019ApJ...881...26V} claimed that the thermal radio component is best estimated using the mixture method as a mixture of H$\alpha$ and 24 $\mu$m calibration. The H$\alpha$ and 24 $\mu$m calibration is from \cite{2007ApJ...666..870C} and \cite{ 2019ApJ...881...26V}:
	\begin{equation} 
		F_{\rm H\alpha,corr}=F_{\rm H\alpha,obs}+ a \cdot F_{24\mu m}.
	 \end{equation}
We adopt a value of a =0.042, there is a linear relationship of 22 $\mu$m flux and 24 $\mu$m flux 
($F_{24\mu m}$ = $1.03F_{22 \mu m}$ from \citealt{2015AJ....150...81W}). 
We used the WISE 22 $\mu$m flux ( $F_{22\mu m}$ $= (1.71 \pm 0.02) \times10^{-11}$ erg s$^{-1}$ cm$^{-2}$ within an elliptical nuclear region of radius $7.5\arcsec\times4.8\arcsec$ from \cite{2013AJ....145....6J} and the observed H$\alpha$ flux $F_{\rm H\alpha,obs}= (3.5 \pm 0.7) \times 10^{-13}$ erg s$^{-1}$ cm$^{-2}$ in the same region to derive the $F_{\rm H\alpha,corr}=(1.1\pm0.3)\times10^{-12}$ erg s$^{-1}$ cm$^{-2}$, thus we estimated the average extinction of $A_{\rm H\alpha}$=$1.3\pm0.4$ within this region. The spatial resolution of the WISE 22 $\mu$m image limits us to get an assessment the extinction-corrected H$\alpha$ flux of these regions, which are smaller than the nuclear region of radius $7.5\arcsec\times4.8\arcsec$. Region T is close to this limit, so we use the average $A_{\rm H\alpha}=1.3\pm0.4$ in region T. \cite{2007ApJ...666..870C} also claimed an uncertainty of 20\% for the coefficient $a$, which we also accounted for to evaluate the uncertainty of $F_{\rm H\alpha,corr}$.

Thus, the thermal radio emission in the region T could be derived from the extinction-corrected H$\alpha$ flux \citep{1992ARA&A..30..575C}, as 
      \begin{gather}\label{eq:thermal}
          \frac{S_T}{\rm mJy}=
            1.25\left(\frac{T_e}{10^4\;\rm K}\right)^{0.59}
                \left(\frac{\nu}{\rm GHz}\right)^{-0.1}
                \left(\frac{F_{\rm H\alpha,corr}}{10^{-12}\;\rm erg\,s^{-1}\,cm^{-2}}\right)
	 \end{gather}
We assumed the electron temperature to be $T_e=10^4$ K, then the derived thermal free-free radio flux densities are 0.66 mJy and 0.64 mJy at 6 GHz and 9 GHz, respectively, and those values are $\sim13\%$ of the total radio emission in the region T of NGC 5792 at 6 GHz and 40\% at 9 GHz. 

We also probed the thermal radio emission of the separate knots on the nuclear ring. However, 
the spatial resolution of the optical and infrared observations limits us to get an assessment the extinction-corrected H$\alpha$ flux of each knot. Instead, we employed the average extinction coefficient of $A_{\rm H\alpha}$=$1.3\pm0.4$ in the $7.5\arcsec\times4.8\arcsec$ radius region, to correct the extinction of the all knots in the region T, and listed the extinction-corrected $F_{\rm H\alpha,corr}$ in Col. 3 of Table \ref{tab:thermal}. We note that the $F_{\rm H\alpha,corr}$ of the six knots A-F potentially were estimated with uncertainties due to different levels of extinction throughout the nuclear rings. As shown in the HST image (Figure \ref{fig:Xband}d), the knots A, B, F and C are apparently more obscured by dust, indicating that the $F_{\rm H\alpha,corr}$ of these knots may be underestimated.

Finally, the thermal radio emission of the separate knots in the region T are assessed by 
Equation (\ref{eq:thermal}). We list the thermal radio emission of the knots in Col. 4 (at 6 GHz) and Col. 6 (at 9 GHz) of Table \ref{tab:thermal}. The fractions of thermal emission to total radio emission are listed in Col.5 (at 6 GHz) and Col.7 (at 9 GHz). All above uncertainties are accounted for using error-propagation techniques.

Furthermore, we probed the variation of $A_{\rm H\alpha}$ inside the nuclear region with the high spatial resolution HST F814W and F475W images, by comparing the observed flux ratio of those two bands to the intrinsic one. The latter one was assessed as 2.9 by modeling the continuous star formation case of constant SFR, by means of the stellar synthesis model STARBURST99 \citep[][]{Leitherer99}.
We found that the $A_{\rm H\alpha}$ of knot A to F varies from 2.0 to 3.6 with lowest values as 2.0 and 2.4 at knot D and E, respectively, and grows as high as 3.6 at knot A. This variation is consistent with what we saw in the multi-color HST image (Figure~\ref{fig:Xband}d), which exhibits a dust lane at the southeast portion. However, the absolute value of $A_{\rm H\alpha}$ relies on the comprehensive investigation of the stellar population inside the nuclear ring, which is beyond the scope of this paper. 


\subsection{Star Formation Rate}    \label{sec:sfr} 
The main goal of this section is to determine the most important parameter, star formation rate (SFR).
Using Equation (2) in \cite{2011ApJ...737...67M} with the extinction-corrected H$\alpha$ emission, we estimated the SFRs of A, B, C, D, E, F, N, R, T,
and core (marked in Figure \ref{fig:Xband}, region R is the ring defined by subtracting the region N from region T, and the core is the point-like source in the center of NGC~5792) as follows:  

        \begin{gather}
        \frac{\rm SFR}{M_{\sun}\,\rm yr^{-1}} = 5.37\times10^{-42} \left( \frac{L_{\rm H\alpha,corr}}{\rm erg~s^{-1}}\right)
	 \end{gather}
The results are listed in Col. 8 of Table \ref{tab:thermal}. 
We assessed the SFR in the nuclear ring (region R) for NGC~5792 of $\sim 0.32$ $M_\sun$ yr$^{-1}$, however the SFR of the total nuclear region (region T, in $\sim1.5$ kpc diamater) is $\sim$ 0.42 $M_\sun$ yr$^{-1}$. It appears that the majority of the star-formation activity is taking place on the ring in the nuclear region. According to \cite{2019ApJ...881...26V}, the SFR of NGC~5792 in $\sim23.7$ kpc diamater is 4.41 $M_\sun$ yr$^{-1}$, which was estimated using a combination of H$\alpha$ and 22 $\mu$m luminosity.  
We also estimated the SFR surface density (SFR$_{\rm SD}$) by dividing the area of the regions. The SFR$_{\rm SD}$ in nuclear region T is 0.46 $M_{\sun}$ yr$^{-1}$ kpc$^{-2}$, the other SFR$_{\rm SD}$ are list Col.9 of Table \ref{tab:thermal}.

\section{Discussion}    \label{sec:dis}
\subsection{The Origin of Non-thermal Radio Emission in the Nuclear Ring} \label{sec:noninring}

We show a nuclear ring with a radius of 5$\arcsec$ (750 pc) in high-resolution 6~GHz and 9~GHz images of NGC~5792 observed by VLA. 
The ring emission dominates the total radio emission in the nuclear region, and its size and orientation in the radio regime is consistent with that in the optical and near infrared bands, as shown in Figure \ref{fig:Xband}. 
Based on the high-resolution data at 6 GHz, we found that the nuclear ring is dominated by non-thermal synchrotron emission. 
The thermal radio continuum emission associated with the star forming process accounts for only $\sim11\%$ of the observed total radio continuum emission. 
The polarization contour also indicates the existence of the non-thermal synchrotron radio emission in the nuclear region, as shown in the top panel of Figure \ref{fig:pol}.

Possible origins of the non-thermal emission include SNe and SNRs related to the star forming process and contribution of AGN activity.
In the case of SNRs as the dominant accelerators of CREs, the excess non-thermal emission should originate from the deaths of massive stars in supernova explosions. Therefore, the non-thermal radio continuum emission traces the star forming process that occurred roughly an average massive star main sequence lifetime ago ($\sim 10$ Myr, \citealt{Leitherer99}), rather than the instantaneous one.
Using Equation (14) of \cite{2011ApJ...737...67M}, we derived a synchrotron-based SFR of 1.2 $M_{\sun}$ yr$^{-1}$ with the non-thermal synchrotron flux of 4.1 mJy at 6 GHz and $\alpha_{1.5-6\rm GHz}\sim-0.7$. This value is roughly three times of the instantaneous SFR of 0.32~$M_{\sun}$~yr$^{-1}$ inferred by using a combination of the H$\alpha$ and 24~$\mu$m emission. This result indicates that the nuclear ring in NGC 5792 underwent more intense star forming activity in the past, and now the star formation in the nuclear ring is in the low state. 
Some numerical simulations found that the ring SFR is closely related to the mass inflow rate \citep[e.g.,][]{2019ApJ...872....5S}. The instantaneous ring SFR may reflect a decreasing of the mass inflow rate.
However, in the nuclear region, the contribution of the AGN to the non-thermal emission cannot be fully ruled out. The existence of an AGN will be discussed in \S\ref{sec:AGN}.

Comparing with the $\alpha_{1.5-6\rm~GHz}\sim-0.70\pm0.15$ \citep{2016MNRAS.456.1723L}, we measured a steeper $\alpha_{6-9\rm~GHz}\sim-2.95\pm0.24$ for the nuclear region.
Other than the potential losing flux density due to the limitation of the LAS as well as synthesized beam at 9 GHz, 
the CREs lose energy by synchrotron, inverse Compton and escape losses can also change the non-thermal spectral index at different frequencies and regions \citep{2007A&A...470..539B}. Especially, the magnetic field is complex with the contribution of AGN in the nuclear region. For example, in the case of NGC~6946's ordered magnetic field, the CREs suffer stronger synchrotron losses hence a rather steep spectrum \citep{2013A&A...552A..19T}. The polarized intensity found in the nuclear region of NGC~5792 is another piece of evidence of this CRE losing energy scenario.


\subsection{Equipartition Magnetic Field Strengths}
Our results suggest that the nuclear region is dominated by non-thermal synchrotron emission. 
Another factor determining the intensity of the non-thermal emission is the magnetic field. Therefore, we estimated the magnetic field from the 6-GHz non-thermal flux density. We assumed the equipartition between the energy densities of the total cosmic rays (dominated by protons) and the total magnetic field. 
We used the revised formula given by \cite{2005AN....326..414B} to calculate the equipartition magnetic fields $B_{\rm eq}$. We also assumed the ratio of the relativistic proton number density to that of the electrons is 100 and a spectral index of $-0.7$, and the path-lengths through the emitting medium along the line of sight are taken to be the same as the maximal widths of the regions in the sky plane. Derived $B_{\rm eq}$ in each region is listed in Col.10 of Table \ref{tab:thermal}. 
The range of $B_{\rm eq}$ in the nuclear ring is from 77 to 88 $\mu$G and a mean of 84 $\mu$G. In the core region of NGC~5792, the $B_{\rm eq}$ goes up to 113 $\mu$G. 

There is a similar nuclear ring in NGC 1097, 
whose radio emission is also dominated by non-thermal emission. The equipartition magnetic field strengths in the ring are changing between 50 and 80 $\mu$G \citep{2005A&A...444..739B}. The mean $B_{\rm eq}$ of NGC 5792 nuclear ring is about 50\% higher than the mean value of $\sim55$ $\mu$G in NGC 1097 \citep{2005A&A...444..739B}. 
\cite{2018NatAs...2...83T} claimed that strong magnetic fields limit the efficiency of massive star formation while fostering enhanced low mass star formation. 
Therefore, the nuclear region of N5792 may be undergoing massive star formation quenching.

\subsection{Star Formation on the Nuclear Ring}
The separation of knots on the nuclear ring of NGC 5792 indicates that star formation is or was concentrated in discrete regions. How do these knots (concentrated star-forming regions) form on the ring?
\cite{2008AJ....135..479B} proposed two models of star formation: the ``popcorn" and ``pearls on a string" models. 
In the former one, star formation occurs in dense clumps that are randomly distributed along a nuclear ring. This type of star formation, presumably caused by the gravitational instability of the ring itself \citep{1994ApJ...425L..73E}, does not produce a systematic gradient in the ages of young star clusters along the azimuthal direction (see also \citealt{2002AJ....123.1411B}). In the latter one, star formation takes place preferentially at the contact points between a ring and the dust lanes. This may happen because the gas clouds with the largest densities are usually placed at the contact points due to orbit crowding \citep{1992ApJ...395L..79K, 1997A&A...319..737R, 1999ApJ...511..157K, 2011ApJ...736..129H}. Since star clusters age as they orbit along the ring, this model consequently predicts a bipolar azimuthal age gradient of star clusters starting from the contact points (see also \cite{2001MNRAS.323..663R, 2006MNRAS.371.1087A, 2008ApJS..174..337M, 2008AJ....135..479B, 2013A&A...551A..81V}). \cite{2008ApJS..174..337M} found that $\sim$50\% of the nuclear rings in their sample galaxies show azimuthal age gradients and that such galaxies have, on average, a larger value of the mean SFR than those without noticeable age gradients.

Observational evidence indicates that the SFR in the nuclear rings of normal barred galaxies spans a wide range $\sim 0.1-10~M_{\sun}~\rm yr^{-1}$ \citep{2008ApJS..174..337M, 2018ApJ...857..116M}. The one of NGC 5792, $\sim0.4~M_{\sun}~\rm yr^{-1}$, is at a low level.
The SFR in the nuclear ring is affected by many factors, for example, the strength of nonaxisymmetric perturbations in galaxies, the inflow rate of the gas, and the magnetic field strength. 
The observations in \citet{2018ApJ...857..116M} support that strongly barred galaxies tend to have low SFRs in their rings \citep[see also in ][]{2008ApJS..174..337M, 2010MNRAS.402.2462C}. Besides that, numerical simulations \citep{2013ApJ...769..100S, 2019ApJ...872....5S} show that the SFR is closely related to the mass inflow rate to the ring rather than the ring mass. For a range of inflow rate $0.125 - 8~M_{\sun}~\rm yr^{-1}$, the SFR is about 80\% of the inflow rate \citep{2021ApJ...914....9M}. The discovery by \citet{2018NatAs...2...83T} shows that the strong magnetic field can decelerate massive star formation and can help the formation of low-mass stars. Further radio observation is needed to address which one is the dominant factor.

    \subsection{Is There an AGN in the Center?}\label{sec:AGN}

We found a potential AGN radio counterpart in the 6 GHz A-configuration map (see Figure \ref{fig:CLband}d), with a flux density of $16\pm4$ $\mu$Jy. 
We fitted the radio core by a Gaussian, and after deconvolving from the synthesized beam we found that the size of the core is less than the FWHM of the synthesized beam in both the major and minor axis directions. 
In other words, the point-like radio core is unresolved in the A-configuration map, which has a resolution of $\sim0.3\arcsec$.
This point-like feature is one of the criteria for determining the presence of an AGN \citep{2019AJ....158...21I}. However, the origin of other kinds of emission components, e.g. SNRs, cannot be ruled out completely.



It is well known that AGNs can have very flat spectra, because AGNs are compact 
and show self-absorbed synchrotron spectra \citep{2010LNP...794..143M}. 
The core region of NGC 5792 shows a non-thermal spectral index of $\alpha_{NT}$, $\le 0.74$ between 6 and 9 GHz. 
On the other hand, according to the $\alpha_{1.5-6\rm GHz}$ of the nuclear region is -0.7, and the flux density of the core is $16~\mu$Jy at 6 GHz , we estimated that the flux density of the core is $\sim 12~\mu$Jy at 9 GHz. This value is too weak to be detected at 9 GHz.

Figure \ref{fig:pol} shows the linear polarization image from the CHANG-ES D-configuration, 6 GHz observations, along with the polarization vectors superimposed. The image reveals two apparent radio lobes on kpc scales, bolstering the AGN hypothesis. \textbf{} We stress that the image has not been corrected for Faraday rotation, and an attempt at reconstructing the image using rotation measure synthesis did not reveal these lobes, possibly because of their faintness.  Nevertheless, the polarized intensity does appear to show evidence of past nuclear activity, similar to what has been found in NGC~2992 \citep{irw17}.

Additionally, the optical \emph{HST} image (see Figure \ref{fig:Xband}d) suggests that the core seems to be obscured by dust in the center of NGC~5792. 
Based on these clues of the point-like feature, the polarized lobes and the feature of the optical image, we suggest that a potential AGN is at the center of NGC 5792. 

The radio luminosity of the core of NGC~5792 is $\sim2.0\times10^{25}$ erg s$^{-1}$ Hz$^{-1}$ at 6 GHz. The nucleus of NGC~5792 was detected as an X-ray source (4XMM J145822.6-010525 in the XMM-Newton Serendipitous Source Catalog\footnote{https://heasarc.gsfc.nasa.gov/W3Browse/xmm-newton/xmmssc.html}) with $L_{0.2-12\rm~keV}\sim2.4-20\times10^{39}$ erg s$^{-1}$.  
 \cite{2014ApJ...789..124D} estimated the mass of the black hole in NGC~5792 as M$_{\bullet}\sim10^{7.17}$ $M_{\sun}$. Thus, the bolometric luminosity of the core is estimated to be $\sim10^{-5} L_{\rm Edd}$.
The radio core of NGC~5792 can thus be interpreted as a low luminosity AGN (LLAGN).

\subsection{Comparison with the Center of Our Galaxy}
The nuclear region of NGC~5792 is reminiscent of the CMZ in our Galaxy. Both of them contain a nuclear ring and an LLAGN. A pair of contact point knots A and B on the nuclear ring of NGC~5792, located opposite each other, are similar to the distribution of the two brightest massive clumps Sgr B2 and Sgr C on the CMZ (see Figure 21 in \citealt{2010ApJ...721..137B}). However, the NGC~5792 nuclear ring is three times larger in size comparing to the CMZ, which has a diameter of $\sim$400~pc \citep{1996ARA&A..34..645M, 2009ApJ...702..178Y}.
In contrast, the stellar mass of NGC~5792 out to a diameter of 72 kpc is 10$^{10.9}$ $M_{\sun}$ \citep{2019PASJ...71S..14S}, which is comparable to the stellar mass of $\sim10^{11}$ $M_{\sun}$  of our Galaxy within a diameter 30 kpc (\citealt{1996A&ARv...7..289M}).

The numerical simulation by \cite{2019ApJ...872....5S} showed that the ring SFR is closely related to the mass inflow rate \textbf{onto} the ring, and suggested that the low current SFR of the CMZ is due to a low mass inflow rate in the near past. The SFR of the CMZ is a level of 0.08-0.15 $M_{\sun}$ yr$^{-1}$ (\citealt{2009ApJ...702..178Y}). 
The star formation of the nuclear ring in NGC 5792 is also in a low state ($\sim0.4$ $M_{\sun}$ yr$^{-1}$), this indicates the low mass inflow rates, however, there are not yet observations to probe this speculation.
Further observations of molecular gas around the nuclear region in NGC 5792 will help to confirm this speculation.

\section{Summary}       \label{sec:sum}
Based on the sub-arcsecond angular resolution radio (at 6 and 9 GHz), near-infrared (\emph{HST}/WFC3 F160W, F814W), and optical (H$\alpha$, \emph{HST}/WFC3 F475W) observations, we investigated the nuclear region of NGC 5792. The study can be summarized as follows,

\begin{enumerate}
\item We detected a well-defined nuclear ring at 6 and 9 GHz based on the VLA A-configuration observations. The ring can also be partially recognized in the \emph{HST} and APO H$\alpha$ images, which suffer significant dust obscuration. 

\item We used the 24 $\mu$m emission to calibrate the dust extinction of the H$\alpha$ emission in the nuclear region, then used the extinction-corrected H$\alpha$ emission to separate thermal and non-thermal emission. We found that the nuclear ring is dominated by the non-thermal synchrotron emission. The excess non-thermal emission may be related to the past star-forming process. 

\item Using the extinction-corrected H$\alpha$ emission, we also estimated the SFRs of regions A, B, C, D, E, F, R, T, and the core. The SFR in the nuclear region, $\sim 0.4 ~M_{\sun}$~yr$^{-1}$, is at a low level among the range that the SFR in the nuclear ring of normal barred galaxies spans. Further radio observations are needed to pin down its dominant cause.
We also assessed the equipartition magnetic field strength $B_{\rm eq}$ of the knots change between 77 and 88 $\mu$G with the synchrotron emission at 6~GHz.
The strong magnetic field may prevent the collapse of gas to form massive stars. 


\item We found that the nuclear radio ring surrounds a faint and compact ($r\sim50$ pc) radio core at the center of NGC~5792 in the 6 GHz image, with a flux density of $16\pm4$ $\mu$Jy and polarized lobes, suggesting that the central source is a putative AGN. 

\end{enumerate} 

\begin{acknowledgments}
Y.Y. and Z.L. acknowledge support by the National Key Research and Development Program of China (2017YFA0402703).
Y.Y. acknowledge sponsored by the Shanghai Sailing Program (19YF1455500).
CJV acknowledges support from the National Science Foundation Graduate
Research Fellowship under Grant No. 127229. RAMW acknowledges support
from the National Science Foundation through Grant AST-1615594.
We acknowledge the referee and reviewers’ constructive and helpful comments for improving our manuscript.
We are grateful to Zhi Li for a discussion of the dynamic origin of the nuclear ring.
\end{acknowledgments}





\begin{deluxetable*}{cccccccc}
\rotate
\tabletypesize{\footnotesize}
\tablecaption{Information of NGC 5792 observations and Images  \label{tab:obs}}
\tablehead{
\colhead{Frequency} & \multicolumn{3}{c}{1.5 GHz (L band)}&\multicolumn{3}{c}{6.0 GHz (C band)}&\colhead{9 GHz (X band)}\\
\cmidrule(r{4pt}){2-4} \cmidrule(r{4pt}){5-7} \cmidrule(r{4pt}){8-8}
\colhead{Array}&\colhead{D$^a$}&\colhead{C$^a$}&\colhead{B$^a$}&\colhead{D$^a$}&\colhead{C$^a$}&\colhead{A$^b$}&\colhead{A$^b$}
}
\startdata 
Date of observations &2011-Dec-30&2012-Mar-30&2011-Apr-05&2011-Dec-10&2012-Feb-04&2015-Jul-28&2015-Jul-09 \\
&&&&&2012-Feb-21&&2015-Jul-11 \\
Total bandwidth (MHz)&512&512&512&2042&2042&2042&2042\\
Obs. time (hr)$^c$&7&8.5&7.5&6.0&4.5&2.3&2.7\\
&&&&&4.5&&2.5 \\
Flux calibrator$^d$&3C286&3C286&3C286&3C286&3C286&3C286&3C286\\
Phase calibrator$^e$&J1505+0326&J1505+0326&J1505+0326&J1505+0326&J1505+0326&J1505+0326&J1505+0326\\
Pol.-leakage calibrator$^f$&J1331+3030	&J1331+3030	&J1331+3030	&J1331+3030	&J1331+3030	&-&-\\
uv weighting$^g$&Briggs&Briggs&Briggs&Briggs&Briggs&Briggs&Briggs\\
\hline
\multicolumn{8}{c}{\textbf{I image}}\\
\hline
Scales for MS-cleaning$^h$ &[0, 10, 20]&[0, 10, 20]&[0, 10, 20, 40]&[0, 10]&[0, 10, 20, 40, 80, 160]&[0, 8, 20]&[0, 3, 6, 15, 60]\\
Synth. beam$^i$ ($\prime\prime$$\times$$\prime\prime$, $\circ$ )&40.1$\times$35.2, -29.5&11.8$\times$11.4, -71.7&3.9$\times$3.4, 48.6&10.3$\times$9.2, 12.7&3.2$\times$2.8, -6.9&0.40$\times$0.26, 46.4&0.24$\times$0.18, -37.9\\
RMS$^j$($\mu$Jy beam$^{-1}$)&45.0&30.0&20.0&18&18&4.0&7.0\\
$S_I$$^k$ (mJy)&$43.5\pm1.2$&$32.7\pm0.4$&$28.8\pm2.0$&$12.8\pm0.3$&$12.8\pm1.1$&$5.9\pm0.1$&$1.6\pm0.1$\\
\hline
\multicolumn{8}{c}{\textbf{P image}}\\
\hline
Synth. beam ($\prime\prime$$\times$$\prime\prime$, $\circ$ )&-&-&-&16.1$\times$15.5, -87.4&-&-&-\\
RMS ($\mu$Jy beam$^{-1}$)&-&-&-&5.0&-&-&-\\
\hline
\enddata
 \vspace{0.3cm}
 \tablecomments{ $^a$ project ID: 10C-119; $^b$ project ID: 15A-400; $^c$ Total observing time before flagging; $^d$ This source was also used as the bandpass calibrator and for determining the absolute position angle for polarization; $^e$ This source is a “primary” calibrator in the sense of its amplitude errors ($<$3\% amplitude closure errors expected) in all arrays and in both bands; $^f$ This zero-polarization calibrator was used to determine the polarization leakage terms; $^g$ Robust = 0 was used in each case, as employed in the CASA clean task; 
 $^h$ Scales used for the multi-scale clean; $^i$Synthesized beam FWHM of major and minor axis, and position angle; $^j$ The root-mean-square error (RMS) was measured manually on each image in an emission-free region; $^k$ Flux densities of the total intensity emission in the nuclear region (nuclear ring and core).}
\end{deluxetable*}

\begin{table*}
\caption{Flux Densities and Spectral Indices in the nuclear region of NGC 5792}
\vspace{0.2cm}
\footnotesize
\begin{tabular}{cccccccc}
\hline
\hline
 Region &Center&Radius &Radius&$S_{6GHz}$&$S_{9GHz}$&Total $\alpha$&Non-thermal $\alpha_{NT}$\\ 
&(J2000)&(arcsec)&(kpc)&(mJy)&(mJy)&($S\propto\nu^{\alpha}$)&$(S\propto\nu^{\alpha_{NT}}$)\\
(1)&(2)&(3)&(4)&(5)&(6)&(7)&(8)\\
\hline
A&224.5960102, -1.091173824		&1.25$\times$0.70	&0.18$\times$0.11	&0.813$\pm$0.033	&0.296$\pm$0.039	&-2.49$\pm$0.34	&-2.58$\pm$0.37\\
B&224.5931713, 1.09131474		&1.33$\times$1.05	&0.20$\times$0.16	&1.337$\pm$0.048	&0.685$\pm$0.050	&-1.65$\pm$0.20	&-1.83$\pm$0.30\\
C&224.5941998, -1.091582332		&0.33			&0.05			&0.105$\pm$0.008	&0.057$\pm$0.013	&-1.49$\pm$0.59	&-1.58$\pm$0.64\\
D&224.5942301, -1.090868643		&0.35			&0.05			&0.095$\pm$0.008	&0.027$\pm$0.014	&-3.06$\pm$1.24	&-4.25$\pm$2.50\\ 
E&224.5947158, -1.090765536		&0.28			&0.04			&0.069$\pm$0.007	&0.021$\pm$0.011	&-2.97$\pm$1.32	&-3.61$\pm$1.93\\
F&224.5943819, -1.091571492		&0.29			&0.04			&0.051$\pm$0.007	&0.028$\pm$0.011	&-1.43$\pm$1.04	&-1.50$\pm$1.11\\
N&224.5945961, -1.091194578		&3.80$\times$0.95	&0.57$\times$0.14	&0.619$\pm$0.046	&0.042$\pm$0.074	&-				&-\\
T&224.5945961, -1.091183738		&7.10$\times$1.80	&1.07$\times$0.27	&5.258$\pm$0.177	&1.592$\pm$0.147	&-2.95$\pm$0.24	&-3.87$\pm$0.80\\
R&224.5945961, -1.091183738		&T-N				&T-N				&4.639$\pm$0.155	&1.550$\pm$0.127	&-2.70$\pm$0.22	&-3.34$\pm$0.58\\
core&224.5945418, -1.091206343	&$<$0.20			&$<$0.03			&0.016$\pm$0.004	&$<$ 0.021	        &$<$0.66			&$<$0.74		  \\
\hline
\end{tabular}
\vspace{0.3cm}

\textbf{Notes}: Regions N and T are ellipses with a PA $\sim84\degr$ of the major axis.
Region R is the ring defined by subtracting region N from the total region T. 
$S_{6GHz}$ and $S_{9GHz}$ are the flux densities of the regions (we convolved the C-band and X-band images to the same beam size;  0.3$\arcsec\times0.3\arcsec$). The core is the magenta cross in the center of NGC~5792. The spectral indices in Column 8 are obtained from the separated non-thermal emission (see \S\ref{sec:thermal} for details).
\label{tab:flux}
\end{table*}

\begin{table*}
\caption{SFRs in the Nucleus of NGC 5792 }
\vspace{0.2cm}
\scriptsize
\begin{tabular}{cccccccccc}
\hline
\hline
Region &$F_{\rm H\alpha . obs}$&$F_{\rm H\alpha . corr}$&$S_{\rm th,6}$&$P_{\rm th,6}$&$S_{\rm th,9}$&$P_{\rm th,9}$ &SFR&SFR$_{\rm SD}$ &$B_{\rm eq}$\\ 
&(10$^{-15}$$\rm erg~ s^{-1} cm^{-2}$)&(10$^{-15}$$\rm erg~s^{-1}cm^{-2}$)&($\mu$Jy)&(\%)&($\mu$Jy)&(\%)&($M_{\sun}~\rm yr^{-1}$)&($M_{\sun}~\rm yr^{-1}~\rm kpc^{-2}$)&($\mu$G)\\
(1)&(2)&(3)&(4)&(5)&(6)&(7)&(8)&(9)&(10)\\

\hline
A	&$5.27\pm1.1$				&$17.5\pm7.3$			&$18.2\pm7.7$		&2	&$17.5\pm7.3$		&6		&$0.011\pm0.005$	&$0.16\pm0.07$		&88\\
B	&$28.5\pm5.8$				&$94.5\pm39.6$		&$98.8\pm41.4$	&7	&$94.9\pm39.8$	&14		&$0.061\pm0.025$	&$0.61\pm0.25$		&86\\
C	&$1.31\pm0.3^{\dag}$		&$4.4\pm1.8$			&$4.6\pm1.9$		&4	&4.4$\pm1.8$		&8		&$0.003\pm0.001$	&$0.36\pm0.15$		&87\\
D	&$3.87\pm0.8^{\dag}$		&$12.8\pm5.4$			&$13.4\pm5.6$		&14	&12.9$\pm5.4$		&47  		&$0.008\pm0.003$	&$0.99\pm0.41$		&80\\
E	&$1.89\pm0.4^{\dag}$		&$6.3\pm2.6$			&$6.5\pm2.7$		&9	&6.3$\pm2.6$		&30		&$0.004\pm0.002$	&$0.72\pm0.30$		&84\\
F	&$0.56\pm0.1^{\dag}$		&$1.9\pm0.8$ 			&$2.0\pm0.8$		&4	&$1.9\pm0.8$		&7		&$0.001\pm0.0005$	&$0.19\pm0.08$		&77\\
N	&$45.3\pm9.1$				&-					&-				&-	&-				&-		&-	&-		&-\\
T	&$191\pm38.2$			&$633\pm266$	&$662\pm277$&13	&$636\pm266$		&40		&$0.415\pm0.170$	&$0.46\pm0.19$		&44\\
R	&$146\pm29.2$			&$485\pm203$			&$505\pm212$		&11	&$485\pm203$		&31		&$0.317\pm0.130$	&$0.49\pm0.20$		&-\\
core	&$0.49\pm0.1^{\dag}$		&$1.6\pm0.7$			&$1.7\pm0.7$		&11	&$1.6\pm0.7$		&8		&$0.001\pm0.0004$	&$<$ $0.40\pm0.16$		&113\\
\hline
\end{tabular}

\vspace{0.3cm}
\footnotesize
\textbf{Notes}: Column 2 lists the observed H$\alpha$ flux, the values marked by by the $\dag$ symbol, whose measured size were smaller than the resolution;
Column 3 lists the extinction-corrected H$\alpha$ flux, obtained by assuming an average extinction of $A_v=1.3$ in the whole nuclear region. 
Column 4 lists the thermal flux densities at 6 GHz; 
Column 5 is the fraction of the thermal emission to total radio emission at 6 GHz; 
Column 6 lists the thermal flux densities at 9 GHz; 
Column 7 is the fraction of the thermal emission to total radio emission at 9 GHz;
Column 8 lists the star formation rate, which were estimated using the extinction-corrected H$\alpha$ emission; 
Column 9 lists the star formation rate surface density; 
Column 10 lists the equipartition magnetic field $B_{\rm eq}$, which were calculated by using the non-thermal 6 GHz flux densities and equipartition energy calculation of \cite{2005AN....326..414B}. We assume the path-lengths in these regions are the same as their maximum diameters, with a constant non-thermal spectral index $-0.7$.

\label{tab:thermal}
\end{table*}

\end{CJK*}
\end{document}